\documentclass[12pt]{article}

\usepackage{amsmath,amsthm,amssymb, amsfonts} 
\usepackage[dvipsnames]{xcolor}
\usepackage[top=1.1in, bottom=1.1in, left=1in, right=1in]{geometry}
\usepackage[onehalfspacing]{setspace}
\usepackage{natbib}
\usepackage{graphicx}
\usepackage{caption}
\usepackage{subcaption} 

\usepackage[colorlinks=true,linkcolor=RawSienna,citecolor=RawSienna,urlcolor=RawSienna,bookmarksopen=true,pdfstartview=FitB]{hyperref}

\newtheorem{theorem}{Theorem}[section]

\usepackage{algorithm}
\usepackage{algpseudocode}

\begin{document}
 
 
\newcommand{\blind}{0} 

\title{Spatially Adaptive Variable Screening in Presurgical fMRI Data Analysis}
\author{Yifei Hu and Xinge Jessie Jeng \\
	\vspace{0.1in}
	Department of Statistics, North Carolina State University \\
    Department of Statistics, North Carolina State University}

\if0\blind
{
	\title{\bf 
		Spatially Adaptive Variable Screening in Presurgical fMRI Data Analysis
		\thanks{Address for correspondence: Xinge Jessie Jeng, Department of Statistics, North Carolina State University, SAS Hall, 2311 Stinson Dr., Raleigh, NC 27695-8203, USA. E-mail: xjjeng@ncsu.edu. 
		}\hspace{.2cm}}
	\author{Yifei Hu and Xinge Jessie Jeng \\ 
	}
	\maketitle
} \fi

\begin{abstract}
Accurate delineation of tumor-adjacent functional brain regions is essential for planning function-preserving neurosurgery. Functional magnetic resonance imaging (fMRI) is increasingly used for presurgical counseling and planning. 
When analyzing presurgical fMRI data, false negatives are more dangerous to the patients than false positives  because patients are more likely to experience significant harm from failing to identify functional regions and subsequently resecting critical tissues. 
In this paper, we propose a novel spatially adaptive variable screening procedure to enable effective control of false negatives while leveraging the spatial structure of fMRI data. 
Compared to existing statistical methods in fMRI data analysis, the new procedure directly control false negatives at a desirable level and is completely data driven. 
The new method is also substantially different from existing false negative control procedures which do not take spatial information into account.   
Numerical examples show that the new method outperforms several state-of-the-art  methods in retaining signal voxels, especially the subtle ones at the boundaries of functional regions, while providing cleaner separation of functional regions from background noise. Such results could be valuable to preserve critical tissues in neurosurgery.

\medskip
	

\end{abstract}

\section{Introduction}
\label{sec:intro}

{Functional Magnetic Resonance Imaging (fMRI) is a powerful noninvasive neuroimaging technique widely utilized to investigate brain areas and networks involved in various cognitive processes \citep{ogawa1990oxygenation, kwong1992dynamic}. It provides a unique window into the dynamic functioning of the human brain, allowing researchers to observe changes in neural activity in response to specific stimuli or tasks.
	An fMRI image consists of a vast array of three-dimensional units called voxels. Each voxel represents a small volume element in the brain, and it serves as the fundamental building block for fMRI data analysis. Within each voxel, fMRI measures the complex interactions between neural activity, local blood flow, and oxygen metabolism. When neurons become active during cognitive processes, they consume more oxygen and energy, leading to increased cerebral blood flow to supply the necessary nutrients and remove waste products. These changes, known as the Blood Oxygenation Level Dependent (BOLD) signal, reflect localized neuronal responses to cognitive processes.}

{In fMRI data analysis, a major goal is to identify the specific voxels that exhibit stimulus-induced signals. These signals indicate regions of the brain that are actively involved in processing the presented stimuli or performing the given tasks. 
	To achieve this goal, various statistical and analytical techniques are employed. One of the most commonly used approaches is the General Linear Model (GLM), which models the BOLD response in each voxel as a function of the experimental design and stimulus timing.
	Another crucial preprocessing step often applied to fMRI data is spatial smoothing. Spatial smoothing involves applying a convolution operation to the fMRI data using a Gaussian kernel. The purpose of smoothing is to reduce noise and enhance the signal-to-noise ratio, making it easier to detect and analyze meaningful brain activations.
	As a result of the smoothing process, the fMRI data becomes spatially correlated, meaning that adjacent voxels tend to have similar signal values.}

{fMRI is commonly employed in presurgical planning to map brain function and identify critical areas that should be preserved during surgery. The tasks used in presurgical planning with fMRI are designed to activate specific brain regions associated with essential functions. Some common tasks include motor tasks to identify regions controlling movement; language tasks to discern language processing areas; sensory tasks to stimulate sensory brain areas; visual tasks for mapping the visual cortex; memory tasks to identify memory-related brain regions; and other cognitive tasks, such as problem-solving, to pinpoint higher-order cognitive functions and corresponding brain areas.
	The central objective of presurgical fMRI is to precisely map functionally relevant brain regions associated with performed tasks. This identification allows surgeons to focus on these specific areas during the surgical procedure, ensuring the preservation of essential cognitive functions while addressing pathological tissues. Therefore, in the realm of presurgical fMRI data analysis, the precision of spatial localization in identifying functionally relevant brain regions is of paramount importance \citep{yoo2004evaluating, haller2009pitfalls}.}

Because the number of voxels in a 3D fMRI image is quite large, typically on the scale of hundreds of thousands, existing statistical methods often perform  multiple testing correction to identify voxels with statistically significant responses. In other words, these methods aim to control the cumulative error of classifying null voxels as functionally relevant when all voxels are tested simultaneously. Commonly used multiple testing methods include the control of familywise error rate (FWER) or false discovery rate (FDR).
Regarding FDR-based approaches in fMRI data analysis, considerable effort has been dedicated to accommodating the spatial structure of fMRI data, see, e.g., \citet{perone2004false, benjamini2007false, schwartzman2008false, zhang2011multiple, shu2015multiple, tansey2018false, cai2021laws}.

Although important progress has been made for spatially aware multiple testing, the existing methods do not directly address the major request in presurgical planning, which is to delineate and protect functional regions. Compared to false positive errors, practitioners are more concerned with false negative errors because
patients are more likely to experience significant harm from mistakenly deeming a region to be functionally uninvolved and subsequently resecting healthy tissues that are vital to the patient's quality of life \citep{loring2002now, durnez2013alternative, liu2016pre, silva2018challenges}. 
Such request motivate us to develop new statistical tools for presurgical fMRI data analysis. 

The false negative control problem considered here is fundamentally different from the  classical power analysis in hypothesis testing. In the framework of hypothesis testing, signal identification hinges on a pre-fixed nominal level of type I error (or some form of cumulative type I errors in multiple testing). Although one can always increase the nominal level of type I error to reduce type II error/false negatives, it is not clear by how much one should increase the nominal level to guarantee a satisfactory control on false negatives. The exact trade-off relationship depends on the signal-to-noise ratio that is unknown in practice \citep{jeng2019variable}. 

In the literature, methods directly addressing false negative control are scarce. 
For the practice of presurginal planning, \citet{liu2016pre} proposed to assign a more severe penalty on false negatives than on false positives in a loss function. The associated weights are determined subjectively based on experts' opinions. 
Data-driven approaches for false negative control have only appeared recently motivated by applications in high-throughput data analyses. For example, \citet{jeng2016rare} developed the AFNC procedure to detect rare variants in genetic association studies, \citet{tony2017optimal} proposed the MDR method to provide a liberal variable selection cutoff in multistage high-throughput studies, \citet{jeng2019efficient} developed AdSMR to address the  heritability gaps for complex traits, and \citet{jeng2023weak} proposed false negative control screening (FNCS) in a general setting with arbitrary covariance dependence.
However, these existing data-driven methods have not considered spatial structures of data, thus would not perform well in fMRI data analysis. This motivates the work presented here.

{In this paper, we present an innovative approach to address false negative control in presurgical fMRI analysis using a data-driven strategy. Since fMRI data contains spatial patterns representing functional brain regions, we leverage the spatially adaptive empirical-Bayes framework from \citet{tansey2018false} to construct a sequential screening procedure.}

{The screening procedure is built upon the effective control of Missed Discovery Rate (MDR), which is defined as the ratio between the expected number of false negatives and the expected number of signal variables \citep{tony2017optimal}. In order to incorporate the spatial structure of the fMRI data, we adopt an empirical-Bayes framework and introduce a new metric named Bayesian MDR (BMDR).    
	The screening procedure unfolds in three steps. First, candidate voxels are sorted based on their estimated posterior probabilities using an empirical-Bayes framework. In the subsequent two steps, we employ a user-specified control level $\beta$($>0$) and execute a sequential screening process to select the smallest subset of voxels with the estimated BMDR below $\beta$. It can be shown that this procedure holds the potential to effectively control MDR at the specified level of $\beta$. We term this novel approach Spatial MDR.}

{In contrast to methods that rely on subjective weighting of false negative and false positive errors within a loss function, Spatial MDR is a data-driven procedure that automatically adapts to the unknown signal-to-noise ratio and the inherent spatial structure of fMRI data. This adaptability enables improved retention of weaker signal voxels, particularly at the boundaries of functional regions. The BMDR metric employed by Spatial MDR can be viewed as a Bayesian version of the MDR metric defined in \cite{tony2017optimal}. However,  in contrast to their approach, we explicitly integrate the spatial structures of the data into our analysis, which enhances the accuracy and reliability of our results for presurgical fMRI data analysis.}

Simulation studies show that (1) Spatial MDR effectively controls false negatives with realized MDR below the nominal level, (2) Spatial MDR is more powerful than several state-of-the-art fMRI analysis methods in identifying signal voxels, especially the subtle ones at the boundaries of functional regions, and (3) Spatial MDR provides cleaner spatial separation of functional regions from background noise compared to the existing false negative control methods. Application to real fMRI data of spatial working memory indicates that Spatial MDR provides more comprehensive discovery for the brain regions related to working memory tasks. Such results could be valuable for function-preserving neurosurgery.

\section{Spatially adaptive false negative control} \label{sec:method}

\subsection{Voxel-specific mixture model}

Let $z_i$ represent measurement intensity at the $i$th voxel arising from a voxel-level statistical model for an experiment.
We assume that $z_i$ follows a voxel-specific mixture model
\begin{equation} \label{def:mixture}
	z_i\sim (1-c_i)\cdot f_0(z_i) + c_i\cdot f_1(z_i),  \qquad i = 1, \ldots, n,
\end{equation}
where $n$ is the total number of voxels, $c_i\in(0, 1)$ is an unknown mixing fraction varying from voxel to voxel, $f_0(\cdot)$ and $f_1(\cdot)$ denote the unknown null ($h_i=0$) and alternative ($h_i=1$) distributions of $z_i$. {Different from the two-groups model commonly used in multiple testing, where $c_i = c$ remains constant for all $i$ \citep{efron2001empirical, muller2006fdr}, Model (\ref{def:mixture}) offers the flexibility of voxel-specific mixing fractions. This flexibility allows for the integration of local spatial information. Model (\ref{def:mixture}) has been applied to analyze fMRI data in \cite{tansey2018false}, where the primary goal is to identify signal voxels while effectively controlling the false discovery rate (FDR).} 

\subsection{Variable screening with MDR control} \label{sec:BMDR}

{Unlike the existing studies that identify signal voxels strong enough to survive multiplicity adjustment in the two-groups or voxel-specific mixture models, we aim to retain not only strong but also relatively weak signal voxels using a principled false negative control strategy.}

{In order to incorporate the spacial structures of fMRI data, we adopt the empirical-Bayes framework used in \cite{tansey2018false} and report the posterior probability of $z_i$ being a signal variable as 
	\begin{equation} \label{def:posterior}
		{w}_i = P(h_i=1 |z_i) =\frac{{c}_i\cdot {f}_1(z_i)}{{c}_i\cdot {f}_1(z_i)+(1-{c}_i)\cdot {f}_0(z_i)},  \qquad i=1,...,n,
	\end{equation}
	where $c_i$, as the prior probability, varies from voxel to voxel, and can be estimated from the data. In this framework, we introduce a new metric, named Bayesian MDR, as follows. }
{For a given decision rule $\delta$, its Bayesian MDR is
	\begin{equation} \label{def:BMDR}
		\text{BMDR}_\delta(\mathbf{z}) = \frac{\sum_{\{i: \delta_i=0\}} w_i}{ \sum_{i=1}^n w_i},
	\end{equation}
	where the numerator is the posterior mean of the number of false negatives (FN) associated with $\delta$, while the denominator is the posterior mean of the total number of signal variables.}
{BMDR can be viewed as a Bayesian version of the MDR metric introduced in \cite{tony2017optimal}. MDR is defined as 
	\begin{equation} \label{def:MDR}
		\text{MDR}_\delta = E(\text{FN}_\delta)/ E(s) = E(\sum_{\{i: \delta_i=0\}} 1(h_i=1))/ E(\sum_{i=1}^n 1(h_i=1)). 
	\end{equation}	
	As MDR essentially quantifies the rate of missed signal variables by the decision rule $\delta$ among all signal variables, a low control level on MDR is associated with a high proportion of signal variables being retained. In their work, \cite{tony2017optimal} developed a MDR control method under the two-groups model setting without taking into account the local spatial information.} 

{Here, our objective is to develop an MDR control method tailored for the voxel-specific mixture model outlined in (\ref{def:mixture}). The proposed BMDR, formulated within the empirical-Bayes framework, facilitates the integration of voxel-level information. We first consider an oracle procedure assuming all the model parameters in (\ref{def:mixture}) are known. The oracle procedure has three steps as follows. }

\begin{itemize}
	\item {
		{\bf Step 1}: Sort the candidate voxels by their $w_i$ values (as in (\ref{def:posterior})) in an decreasing order such that ${w}_{(1)} \ge {w}_{(2)} \ge ... \ge {w}_{(n)}$. Consider a decision rule of the form
		\begin{equation} \label{def:delta}
			\delta(\mathbf{w},j) = (1(w_1 \ge w_{(j)}),\ldots, 1(w_n \ge w_{(j)} )).
	\end{equation}}
	\item {{\bf Step 2}:  Implement a user-specified MDR control level $\beta(>0)$ and calculate 
		\begin{equation} \label{def:j*}
			j^*=\min{\{j: \text{BMDR}_{\delta(\mathbf{w},j)} (\mathbf{z}) < \beta \}},
		\end{equation}
		where $\text{BMDR}_{\delta(\mathbf{w},j)} (\mathbf{z})$ is as defined in (\ref{def:BMDR}). }
	\item {
		{\bf Step 3}: Select voxels based on the decision rule $\delta(\mathbf{w},j^*)$. }
\end{itemize}

{It can be seen that for the sequential decision rule $\delta(\mathbf{w},j), j=1, \ldots, n$,  BMDR$_{\delta(\mathbf{w},j)}$ is non-increasing with respective to $j$. Consequently, $j^*$ corresponds to the initial instance when BMDR${\delta(\mathbf{w},j)}$ becomes less than $\beta$. The set of voxels selected by $\delta(\mathbf{w},j^*)$ is the smallest set of voxels with BMDR$_{\delta(\mathbf{w},j^*)} < \beta$. The following theorem formalizes the MDR control property inherent in this procedure for fully-specified model (\ref{def:mixture}). 
	\begin{theorem} \label{thm:MDR control}
		Consider the voxel-specific mixture model in (\ref{def:mixture}). Given a user-specified control level $\beta (>0)$ on MDR, the decision rule $\delta(\mathbf{w},j^*)$, constructed according to (\ref{def:delta}) - (\ref{def:j*}), ensures $\text{MDR}_{\delta(\mathbf{w},j^*)} \le \beta$.   
\end{theorem}}

{Like other methods employing pre-fixed nominal levels for error control, the proposed method requires a pre-fixed $\beta$ level for MDR control. By choosing a control level sufficiently low, such as 0.05 or 0.1, the method enables effective retention of weak signal voxels, particularly those located at the boundaries of functional regions. Moreover, the procedure strives to select the smallest subset of voxels under MDR control, thereby preventing the inclusion of an excessive number of noise voxels in the background.}

\subsection{Spatially adaptive MDR control procedure}

In the empirical-Bayes framework that generate the posterior probabilities ($w_i, i=1, \ldots, n$), the parameters in (\ref{def:mixture}), if unknown, can be estimated from the data.
Specifically, we have the estimated posterior probabilities
\begin{equation} \label{def:est_posterior}
	\hat{w}_i= \frac{\hat{c}_i\cdot \hat{f}_1(z_i)}{\hat{c}_i\cdot \hat{f}_1(z_i)+(1-\hat{c}_i)\cdot \hat{f}_0(z_i)}, \qquad i=1, \ldots, n,
\end{equation}
where $\hat c_i$ is the estimated prior probability that $z_i$ measures the intensity of a signal voxel, $\hat f_0$ and $\hat f_1$ are the estimated noise and signal density functions of $z_i$. 

To obtain the estimated prior $\hat c_i$, we employ the locally adaptive estimator from \citet{tansey2018false} to
accommodate the spatial structure of fMRI data as follows. Assume
\begin{equation}
	c_i=\frac{e^{\gamma_i}}{1+e^{\gamma_i}}.
\end{equation}
Thus, $e^{\gamma_i}$ is the prior odds that the $i$th voxel has a signal, and $\gamma_i$ is the unknown log odds. The unknown $\gamma_1, \ldots, \gamma_n$ can be estimated by solving a non-standard high-dimensional optimization problem based on a graph. 

Assume that each voxel is a node in an undirected graph $\mathcal{G}$ with edge set $\mathcal{E}$. To enforce spatial smoothness in the estimation of $\gamma_1, \ldots, \gamma_n$, penalization on the pairwise differences $|\gamma_i-\gamma_j|$ over the graph $\mathcal{G}$ is imposed as follows:  
\begin{equation} \label{eq:optimize}
	\hat{\mathbf{\gamma}} = \arg\min_{\mathbf{\gamma} \in \mathbb{R}^n} l(\mathbf{\gamma}) +  \lambda \sum_{(i, j)\in\mathcal{E}} |\gamma_i-\gamma_j|,
\end{equation}
where $l(\mathbf{\gamma})$ is the negative log likelihood function with fixed $f_0$ and $f_1$, i.e. 
\[
l(\mathbf{\gamma})=-\sum_{i=1}^{n}\log{\left[\left( \frac{e^{\gamma_i}}{1+e^{\gamma_i}}\right)f_1(z_i)+\left(1- \frac{e^{\gamma_i}}{1+e^{\gamma_i}}\right)f_0(z_i)\right] }.
\]
Because the $L_1$ penalty in (\ref{eq:optimize}) encourages similar $\hat \gamma_i$ values across edges of the graph, the solution $\hat \gamma$ will partition nodes of the graph into regions where $\hat \gamma_i$ are locally constant or approximately equal.  
Therefore, the estimated prior ($\hat c_i = e^{\hat \gamma_i} / (1+e^{\hat \gamma_i}))$ inherits the spatial information and is implemented to obtain the posterior probability $\hat{w}_i$ as in  (\ref{def:est_posterior}). Moreover, the estimated density functions $\hat f_0$ and $\hat f_1$ can be obtained by existing methods ({references...}). 

The spatially aware posterior probability $\hat{w}_i$ are implemented in our sequential screening procedure, as described in Section \ref{sec:BMDR}, to generate the final decision rule $\delta(\mathbf{\hat w}, j^*)$. We refer to the entire procedure as Spatial MDR and advocate its application in the analysis of presurgical fMRI data. 

Spatial MDR is computationally feasible for high-resolution fMRI data. \citet{tansey2018false} provides detailed discussions on how to solve the graph-fused optimization problem in (\ref{eq:optimize}) by an efficient augmented-Lagrangian algorithm and existing methods to derive  $\hat f_0$  and $\hat f_1$ in (\ref{def:est_posterior}). 
For the $128 \times 128$ image data studied in Section \ref{sec:application},  the computing time of Spatial MDR is about $77$ seconds using a MacBook Pro with CPU 2.9 GHz Core i7 and 16 GB memory. The method has been implemented in a software developed in Python. The software, together with a sample
input data set and complete documentation, are publicly accessible at https://github.com/yifeihu93/smdr.

\section{Simulation Studies} \label{sec:simulation}

\subsection{Simulation Setup}

We generate a grid graph with $128\times128$ voxels, which contains two overlapping round signal areas with radius 15 and 20, respectively. The signal region has in total 1686 voxels as shown in the first plot of Figure \ref{fig:sim_plot0}. For each voxel, $z$ value is generated from the two-groups model:

\begin{equation} \label{eq:simulation}
	z_i\sim c_i\cdot N(\theta_i, 1) +(1-c_i)\cdot N(0, 1). 
\end{equation}
where $\theta_i$ represents the intensity level of signal distribution.  We set $c_i=1$ for $i$ within the signal region and $c_i=0$ or $0.05$ for $i$ outside the signal region. We consider random $\theta_i$ and assume two different distributions for $\theta_i$: $\theta_i \sim 0.5N(-2, 1)+0.5N(2, 1)$ versus $\theta_i \sim N(0, 3)$.  The setting with $\theta_i \sim N(0, 3)$ is referred to as a ``poorly separated" case in  \citet{tansey2018false} because the null and signal components in (\ref{eq:simulation}) have the same mean/mode.  
These scenarios are summarized as follows.  

\begin{itemize}
	\item Well separated:  $\theta_i \sim 0.5N(-2, 1)+0.5N(2, 1)$. 
	\item Poorly separated:  $\theta_i \sim N(0, 3)$.
	\item Pure background: $c_i=0$ for $i$ outside the signal region. 
	\item Noisy background: $c_i=0.05$  for $i$ outside the signal region. 
\end{itemize}

\subsection{Comparison to spatially aware multiple testing} \label{sec:sim_multiple}

We benchmark Spatial MDR (SMDR) against three multiple testing methods that have been applied to fMRI data analysis: (a) BH-FDR \citep{benjamini1995controlling} is a generic procedure that can control FDR under arbitrary dependence; (b) $\text{FDR}_L$ \citep{zhang2011multiple} utilizes spatially smoothed $p$-values to improve power for imaging data analysis; and (c) FDR smoothing (FDR$_S$, \cite{tansey2018false}) performs simultaneous clustering and FDR control via spatially adaptive local false discovery rate (Lfdr$_i = 1-\hat w_i$, with $\hat w_i$ as in (\ref{def:posterior})). The nominal FDR control levels for all three methods are set at $\alpha=0.05$. The nominal MDR control level of SMDR is set at $\beta=0.1$. 

Table \ref{tab:comp_testing} presents the empirical MDR and FDR 
of the four methods in different simulation settings. It shows that (1) the multiple testing methods have low empirical FDR values, generally less than their nominal FDR level of 0.05; (2) FDR$_L$ and FDR$_S$ improve upon BH-FDR by taking spatial information into account, thus their empirical MDR values are lower but not under the nominal MDR level of 0.1; and (3) SMDR has empirical MDR values less than the nominal level of 0.1 in all the settings.  

Results here demonstrate different utilities of the FDR related methods and SMDR. The latter could be more helpful for effective false negative control.  Although a naive approach can be proposed to increase the power of FDR methods by increasing their nominal FDR levels, in practice, it is difficult to provide an FDR level a priori to achieve the nominal MDR level, as that eventually depends on the unknown signal-to-noise ratio.  

\begin{table}[h]
	\renewcommand\arraystretch{1.2}
	\caption{Comparison with FDR methods. The empirical MDR and FDR values of different methods are calculated from 100 replications. Standard deviations are in parentheses. The nominal FDR control levels of BH-FDR, FDR$_L$, and FDR$_S$ are set at $\alpha=0.05$. The nominal MDR control level of SMDR is set at $\beta=0.1$.} \label{tab:comp_testing}	
	\centering
	\setlength{\tabcolsep}{2mm}{
		\begin{tabular}{ccccc}
			\hline
			Model & Background & Method & MDR & FDR \\
			\hline
			\hline
			&     & BH-FDR & 0.818 (0.013) & 0.044 (0.012) \\
			&  Pure   & FDR$_L$   & 0.409 (0.032) & 0.056 (0.013) \\
			&     & FDR$_S$   & 0.336 (0.047) & 0.008 (0.003) \\
			Well separated &     & \textbf{SMDR}   & \textbf{0.027 (0.013)} & 0.107 (0.084) \\
			\cline{2-5}
			&      & BH-FDR    & 0.811 (0.012) & 0.043 (0.011) \\
			&  Noisy    & FDR$_L$  & 0.448 (0.032) & 0.065 (0.013) \\
			&      & FDR$_S$  & 0.395 (0.045) & 0.009 (0.003) \\
			&      & \textbf{SMDR}  & \textbf{0.055 (0.010)} & 0.286 (0.100) \\
			\hline
			&     & BH-FDR     & 0.672 (0.014) & 0.044 (0.009) \\
			&  Pure  & FDR$_L$   & 0.377 (0.028) & 0.056 (0.012) \\
			&     & FDR$_S$   & 0.359 (0.041) & 0.007 (0.003) \\
			Poorly separated &     & \textbf{SMDR}   & \textbf{0.065 (0.031)} & 0.029 (0.028) \\
			\cline{2-5}
			&      & BH-FDR    & 0.668 (0.014) & 0.044 (0.009) \\
			&  Noisy    & FDR$_L$  & 0.424 (0.025) & 0.064 (0.013) \\
			&      & FDR$_S$  & 0.403 (0.038) & 0.009 (0.004) \\
			&      & \textbf{SMDR}  & \textbf{0.067 (0.020)} & 0.158 (0.095) \\
			\hline
	\end{tabular} }		
\end{table}

\subsection{Comparison to other false negative control methods} \label{sec:sim_FN}

In this section, we compare SMDR with two existing false negative control methods: the original MDR \citep{tony2017optimal} and the AFNC procedure developed in \cite{jeng2016rare}.
All three methods are data-driven and require a pre-fixed control level on false negatives. However, MDR and AFNC were developed under independence and do not take spatial dependence into consideration. 

We apply the three false negative control methods with their nominal control levels set at 0.1.
Their performances are measured in empirical MDR, FDR, and  
Fowlkes-Mallows (FM) index \citep{fowlkes1983method, tharwat2020classification}. 
The FM-index has been used to assess classification methods by summarizing FNP $(\# \text{false negatives}/\#\text{signals})$ and FDP $(\#\text{false positives}/\#\text{selected cases})$ as follows:
\begin{equation*}
	\text{FM-index}=\sqrt{(1-\text{FNP})\times(1-\text{FDP})}.
\end{equation*}
Higher FM-index indicates better classification of signal and noise cases. This measure is suitable here because all three methods serve to control FNP or its mean value, MDR. If the controlling purposes are achieved by the methods under comparison, then the one with a higher FM-index is more efficient. Note that this measure is not appropriate to compare the methods in Section \ref{sec:sim_multiple}, which serve for very different controlling priorities.

Table \ref{tab:comp_FN} summarizes results over various simulation settings, where the advantage of SMDR is clearly demonstrated. First,  only SMDR has empirical MDR controlled under the nominal level of 0.1 in all the scenarios, while the other two methods failed to do so. Secondly, the empirical FDR of SMDR is much lower than those of MDR and AFNC, indicating less false positives for SMDR. Thirdly, the FM-index of SMDR is the highest among the three methods, indicating the best overall performance in separating signal and noise voxels.

\begin{table}[h]
	\renewcommand\arraystretch{1.2}
	\caption{Comparison with other false negative control methods. The empirical MDR, FDR, and FM-index are calculated from 100 replications. Standard deviations are in parentheses. The nominal levels of all three methods are set at $\beta=0.1$.} \label{tab:comp_FN}
	\centering
	\setlength{\tabcolsep}{2mm}{
		\begin{tabular}{cccccc}
			\hline
			Model & Background & Method & MDR & FDR & FM-index\\
			\hline
			\hline
			&     & MDR    & 0.522 (0.019) & 0.407 (0.026) & 0.532 (0.012)\\
			&  Pure   & AFNC   & 0.219 (0.251) & 0.697 (0.237) & 0.409 (0.102)\\
			Well separated &     & \textbf{SMDR}   & \textbf{0.027 (0.013)} & \textbf{0.107 (0.084)} & \textbf{0.931 (0.041)}\\
			\cline{2-6}
			&      & MDR   & 0.519 (0.018) & 0.383 (0.027) & 0.544 (0.011)\\
			&  Noisy    & AFNC  & 0.211 (0.246) & 0.690 (0.238) & 0.420 (0.100)\\
			&      & \textbf{SMDR}  & \textbf{0.055 (0.010)} & \textbf{0.286 (0.100)} & \textbf{0.819 (0.054)}\\
			\hline
			&     & MDR    & 0.509 (0.018) & 0.327 (0.041) & 0.574 (0.015)\\
			&  Pure   & AFNC   & 0.206 (0.240) & 0.678 (0.270) & 0.420 (0.120)\\
			Poorly separated &     & \textbf{SMDR}   & \textbf{0.065 (0.031)} & \textbf{0.029 (0.028)} & \textbf{0.953 (0.014)}\\
			\cline{2-6}
			&     & MDR   & 0.511 (0.022) & 0.305 (0.037) & 0.582 (0.011)\\
			&   Noisy   & AFNC  & 0.209 (0.238) & 0.664 (0.270) & 0.434 (0.117)\\
			&      & \textbf{SMDR}  & \textbf{0.067 (0.020)} & \textbf{0.158 (0.095)} & \textbf{0.884 (0.045)}\\
			\hline
	\end{tabular}}
\end{table}

\subsection{Results presented in 2-D grid graphs}

\begin{figure}[h]
	\centering
	\includegraphics[height = 0.5\textwidth] {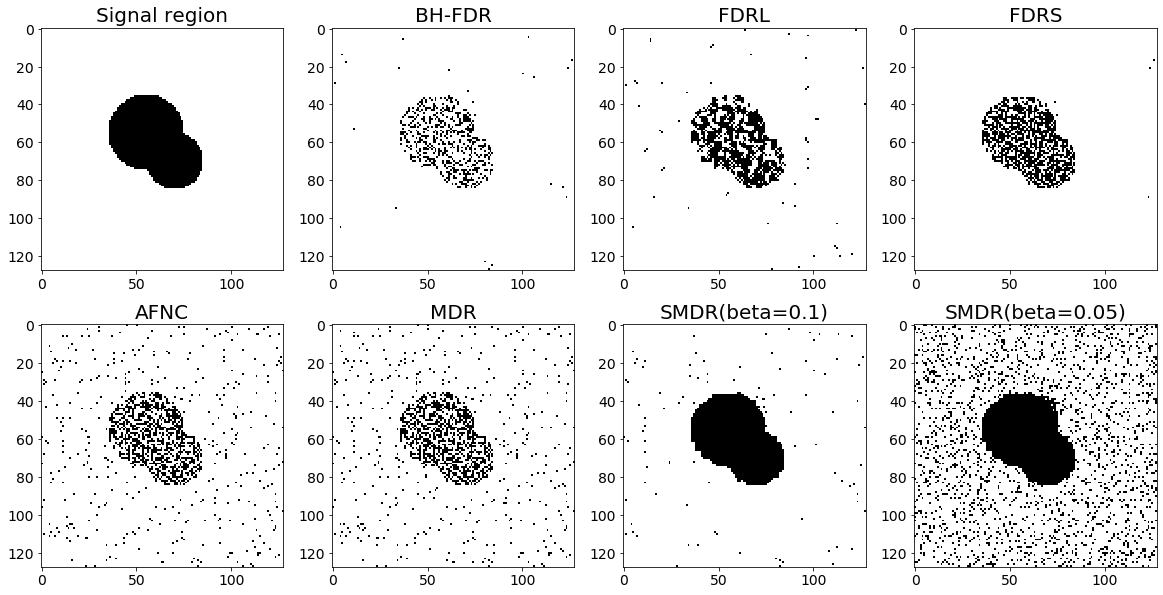}
	\caption{Results from a single trial generated with strong signal effects ($\theta \sim N(0,3))$. The first plot illustrates the signal region. The remaining seven plots illustrate signal voxels identified by different approaches. } \label{fig:sim_plot0} 
\end{figure}

We utilize 2-D grid graphs to illustrate the results of all six methods investigated in Sections \ref{sec:sim_multiple} and \ref{sec:sim_FN}. Figures \ref{fig:sim_plot0} 
presents the 
results from a single trial generated under the setting with pure background and heteroscedastic mixtures ($\theta \sim N(0, 3)$). 

The first plot in Figure \ref{fig:sim_plot0} shows the signal region. The other plots in the top row present the results of the three multiple testing methods: BH-FDR, FDR$_L$, and FDR$_S$. Among these three methods, 
BH-FDR selects the least amount of voxels and misses many voxels in the signal region. 
FDR$_L$ and FDR$_S$ identify more voxels in the signal region without substantially increasing false positives. FDR$_S$ seems to outperform FDR$_L$ in selecting less noise voxels in the background.

Plots in the bottom row of Figure \ref{fig:sim_plot0} are for the false negative control methods AFNC($\beta=0.1$), MDR($\beta=0.1$) and SMDR($\beta=0.1$ and $0.05$). Among these three methods with the same nominal levels, SMDR performs the best in identifying more signal voxels and less noise voxels. In the last plot, SMDR with a lower $\beta$ value selects more signal voxels. Because of the well-known trade-off between false negative and false positive errors, more stringent false negative control also results in more noise voxels being selected. Noticing that the selected noise voxels are mostly scattered in the background, a moderate amount of them are not likely to affect the identification of signal regions.  

Overall, the  proposed SMDR selects significantly more signal voxels than all the other methods. By leveraging the spatial structure of fMRI data, SMDR provides cleaner separation of signal regions from background noise and almost fully recover the signal region.

\subsection{Performance of the spatially adaptive estimator in SMDR}

As a part of Spatial MDR, the spatially adaptive estimator constructed in Step 1 provides an estimate for the total number of signal voxels.  
We compare the performance of the new estimator with that of the estimator implemented in the independent MDR method, which was developed in \citet{jin2007estimating} to estimate the number of signals among independent variables.

Table \ref{tab:s_hat} summarizes the ratios between the estimated values ($\hat s$) and the true value ($s$) of the two methods in the different simulation scenarios. It shows that our spatially adaptive estimator outperforms the estimator in \cite{jin2007estimating} with its $\hat s/s$ ratios much closer to $1$. 

\begin{table}[h]
	\renewcommand\arraystretch{1}
	\caption{Comparison of two estimators for $s$: the one developed under independence and implemented in the original MDR method (denoted as JC) versus the new estimator provided by Spatial MDR (denoted as SMDR). Mean values of $\hat s/s$ are calculated from 100 replications. Standard deviations are presented in parentheses. } \label{tab:s_hat}
	\begin{center}
		\setlength{\tabcolsep}{6mm}{
			\begin{tabular}{cccc}
				\hline
				Model & Background & Method & $\hat s / s$\\
				\hline
				\hline
				& Pure 
				& JC     & 0.85 (0.15) \\
				Well separated     & & SMDR   & \textbf{0.94 (0.04)} \\
				\cline{2-4}
				& Noisy 
				& JC    & 0.85 (0.14) \\
				& & SMDR   & \textbf{0.92 (0.04)} \\
				\hline
				& Pure 
				& JC    & 0.77 (0.14) \\
				Poorly separated     & & SMDR  & \textbf{0.86 (0.04)} \\
				\cline{2-4}
				& Noisy 
				& JC    & 0.78 (0.14) \\
				& & SMDR   & \textbf{0.85 (0.05)} \\
				\hline
		\end{tabular}}
	\end{center}
\end{table}

\section{Data Application} \label{sec:application}

\begin{figure}[h]
	\centering
	\includegraphics[height = 0.85\textwidth]{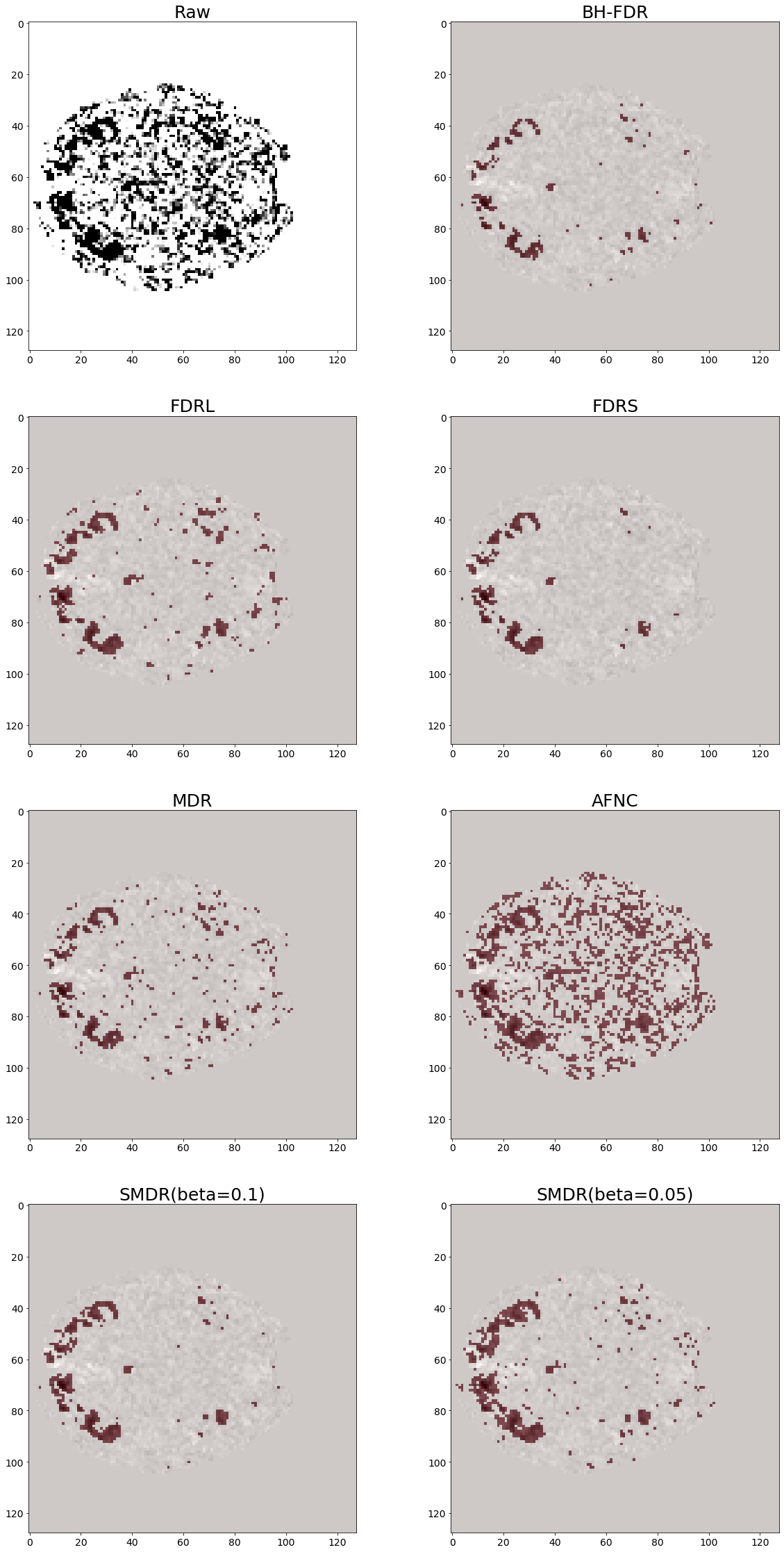}
	\caption{Results of an fMRI experiment on spatial working memory. The first plot illustrates the raw $z$ scores.  The other plots illustrate discoveries made by different methods. } \label{fig:real}
\end{figure}

We apply Spatial MDR to analyze data from an fMRI experiment on spatial working memory.
The full 3-D image of the experiment has  128 $\times$ 128 $\times$ 75 voxels. A single 128 $\times$ 128 horizontal slice image is presented in the first plot of Figure \ref{fig:real}, where darker shade represents higher absolute value of the $z$ score. 
Details about the experiment and the process to generate $z$ scores can the found in \citet{tansey2018false}. 
Other plots in Figure \ref{fig:real} present the discoveries of different methods. 

It can be seen that, similar to the results observed in simulation studies, BH-FDR selects the least amount of voxels; FDR$_L$ and FDR$_S$ select more voxels than BH-FDR and reveal clustered signal regions; AFNC and the original MDR also select more voxels than BH-FDR while including many noise voxels in the background. Among these existing methods, FDR$_S$ seems to perform the best in distinguishing signal regions from noise. 

Compared to existing methods, SMDR demonstrated in the last row of Figure \ref{fig:real} identifies fuller signal region without including excessive noise voxels. Its adaptivity to varying $\beta$ level allows additional flexibility to meet users' needs.

\begin{figure}[h]
	\centering
	\includegraphics[height = 0.42\textwidth]{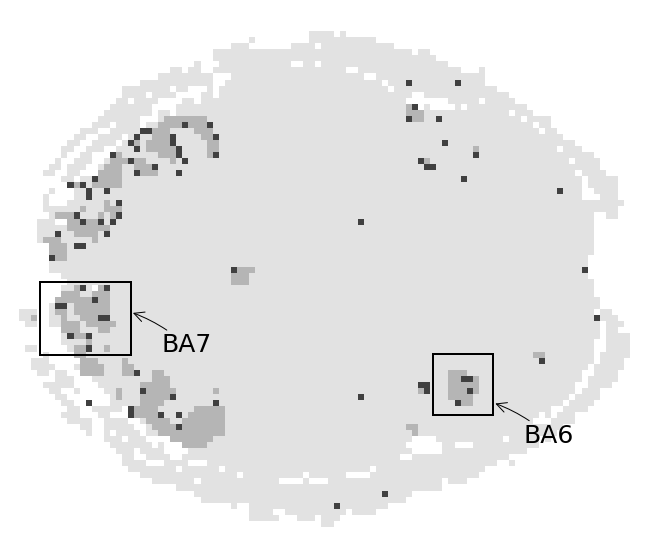}
	\includegraphics[height = 0.42\textwidth]{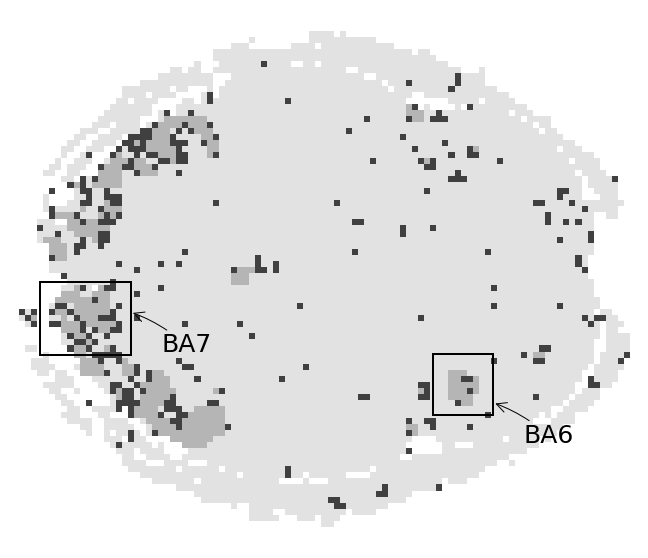}
	\caption{Left plot shows the difference between SMDR$_{\beta=0.1}$ and FDR$_S$. Right plot shows the difference between SMDR$_{\beta=0.05}$ and FDR$_S$. Grey areas are common discoveries of the two methods. Black areas are the additional voxels only selected by SMDR. BA6 and BA7 are two functionally relevant regions for working memory tasks.} \label{fig:data_boundary}
\end{figure}

We specifically compare the results of SMDR and FDR$_S$ in Figure \ref{fig:data_boundary}, where it shows that SMDR selects all the voxels that are selected by FDR$_S$. Those additional voxels only selected by SMDR mostly locate at the boundaries of signal clusters, where signal effects are relatively weak. Specifically, there are two regions called Brodmann area (BA) 6 at the bilateral premotor cortex and BA 7 at the bilateral and medial posterior parietal cortex, which are believed to be related to working memory tasks \citep{owen2005n}. SMDR seems to provide better recovery for BA6 and BA7. 
Overall, SMDR demonstrates substantially better power in identifying signal voxels at the boundaries of functional regions, which could be helpful for planning function-preserving neurosurgery.

\section{Conclusion and discussion} \label{sec:discuss}

This paper introduces a novel variable screening method tailored for presurgical fMRI data analysis. The proposed approach, Spatial MDR, builds upon the newly developed BMDR metric for false negative control, within the context of a voxel-specific mixture model setting. Spatial MDR addresses two critical requirements in planning function-preserving neurosurgery. "Firstly, it provides a principled, data-driven approach to effectively control false negative errors, allowing for the accurate identification and protection of functional regions from damage during surgery. Secondly, Spatial MDR capitalizes on the spatial structure inherent in fMRI data, facilitating improved separation between functional regions and noisy background.

{Spatial MDR offers a theoretical guarantee for MDR control in a fully specified voxel-specific mixture model. When dealing with cases where model components require estimation, we utilize a plug-in procedure that leverages spatially adaptive estimators developed in \cite{tansey2018false}. The derivation of these estimates involves solving a non-standard high-dimensional optimization problem based on a graph, making the theoretical proof of their consistency intricate and extensive, and it is therefore deferred to future research.
	We conduct extensive numerical analyses, demonstrating the validity and superiority of Spatial MDR over other state-of-the-art methods in retaining signal voxels through MDR control.}

{The utilization of Spatial MDR requires the specification of a nominal control level $\beta$ on MDR. A lower $\beta$ indicates a more stringent control of MDR, leading to the selection of more signal voxels. However, this heightened stringency comes at the cost of potentially increasing false positives. Despite this trade-off, Spatial MDR is fundamentally different from methods employing arbitrary thresholding on voxel-level test statistics. This distinction emerges due to the intrinsic association of the $\beta$ value with the MDR level, giving rise to a principled thresholding approach. This distinction becomes particularly evident in a specific scenario where all signal variables precede noise variables in significance. In such cases, Spatial MDR, operating as a sequential screening procedure, can exclusively identifies signal variables even with $\beta = 0$. In contrast, an unprincipled thresholding strategy may still include excessive noise variables in its selection.}

In our numerical examples provided in Section \ref{sec:simulation} and \ref{sec:application}, we set $\beta$ at the low levels of $0.1$ or $0.05$, as these values are commonly used nominal levels in hypothesis testing, even though $\beta$ is primarily designed for addressing false negative errors rather than false positive errors.
For real-world applications, we strongly advise practitioners to report the chosen $\beta$ level alongside their results to facilitate meaningful cross-study comparisons.

{In the practical context of presurgical planning, we recommend practitioners to explore multiple levels of MDR control, taking into consideration the tumor's location. When the tumor is in proximity to a subtle or small active region, it is advisable to choose a smaller $\beta$ value. However, since a more stringent control level may lead to the inclusion of both weak signal voxels and noise voxels, it becomes crucial to distinguish between these two voxel types to some extent. Fortunately, in presurgical applications, weak signal voxels often appear at the boundaries of functional regions, while noise voxels are scattered throughout the background.
	We suggest that practitioners can experiment by gradually decreasing $\beta$ until the additionally selected voxels are predominantly scattered in the background, adding little significant contribution to the identification of fuller active regions. During this stage, it is essential to leverage the expertise of practitioners to make informed and nuanced final decisions. Following this, the identified regions can undergo thorough verification and refinement during the planning phase.}

Last but not least, we acknowledge that apart from the empirical-Bayes framework, several other methods have been proposed to handle spatial structures in image data analyses. Notably, locally adaptive $p$-values have been developed to incorporate valuable local patterns \citep{zhang2011multiple, cai2021laws}. Exploring false negative control strategies for various spatially adaptive measures presents an intriguing avenue to enhance weak signal retention in image data analyses. 

\section*{Supporting Information}

Software developed in Python, together with a sample
input data set and complete documentation is available
at Github: https://github.com/yifeihu93/smdr. 

\section*{Appendix} 

\subsection{Proof of Theorem \ref{thm:MDR control}}

{
	For notation simplicity, let $\delta^* = \delta(\mathbf{w},j^*)$. Then, by (\ref{def:MDR}),  it is enough to show that
	\[
	E(\text{FN}_{\delta^*}) \le \beta \cdot E(s). 
	\]
	First, consider $E(\text{FN}_{\delta^*}|\delta^*)$. We have
	\begin{eqnarray} \label{1-1}
		E(\text{FN}_{\delta^*}|\delta^*) & = & E(\sum_{\{i: \delta^*_i=0\}} 1(h_i=1) | \delta^*) = \sum_{\{i: \delta^*_i=0\}} P(h_i=1) \nonumber\\
		& = & \sum_{\{i: \delta^*_i=0\}} E\left(P(h_i=1|z_i)\right) =  E\left(\sum_{\{i: \delta^*_i=0\}} P(h_i=1|z_i)\right) = E \left(\sum_{\{i: \delta^*_i=0\}} w_i\right)
	\end{eqnarray}
	where the second equality is due to the fact that the event $\{h_i=1\}$ does not depend on the decision rule $\delta^*$, and the last equality is by the definition of $w_i$ in (\ref{def:posterior}).}

{On the other hand, the construction of $\delta^*$ implies BMDR$_{\delta^*} < \beta$, which, by (\ref{def:BMDR}),  implies 
	$\sum_{\{i: \delta^*_i=0\}} w_i < \beta \cdot \sum_{i=1}^n w_i$ almost surely. Then we have 
	\begin{equation} \label{1-2}
		E \left(\sum_{\{i: \delta^*_i=0\}} w_i\right) < \beta \cdot E(\sum_{i=1}^n w_i) = \beta \cdot \sum_{i=1}^n P(h_i=1) = \beta \cdot E(s). 
	\end{equation}
	Combining (\ref{1-1}) and (\ref{1-2}) gives 
	\[
	E(\text{FN}_{\delta^*}) = E\left(E(\text{FN}_{\delta^*}|\delta^*)\right) \le \beta \cdot E(s),
	\]
	which concludes the proof. }

\bibliographystyle{chicago}
\bibliography{refs.bib}

 
\end{document}